\documentstyle[preprint,eqsecnum,aps]{revtex}

\begin{document} 
\tighten
\draft
\newcommand{\MF}{{\large{\manual META}\-{\manual FONT}}}
\newcommand{\manual}{rm}        
\newcommand\bs{\char '134 }     
%
\newcommand{\Vec}[1]{\mbox{\boldmath$#1$}}   
%

\title{Inclusive $K^+$ and exclusive $K^+Y$ photoproduction
on the deuteron: $\Lambda$- and $\Sigma$-threshold
phenomena}

\author{H. Yamamura, K. Miyagawa}

\address{
Department of Applied Physics, Okayama University of Science\\
1-1 Ridai-cho, Okayama 700, Japan
}

\author{T. Mart}

\address{Jurusan, Fisika, FMIPA, Universitas Indonesia,
Depok 16424, Indonesia}

\author{C. Bennhold}
\address{Center for Nuclear Studies, Department of Physics,
The George Washington University, Washington, D.C. 20052,
USA}

\author{W. Gl\"ockle}
\address{
Institut f\"ur Theoretische Physik II, Ruhr-Universit\"at
Bochum, D-44780 Bochum, Germany
}

\maketitle

\begin{abstract}
Inclusive $K^+$ and exclusive $K^+Y$ photoproduction
on the deuteron are investigated theoretically.
Modern hyperon-nucleon forces and a recently updated
kaon photoproduction operator for the
$\gamma +N\rightarrow K^++Y$ process are used. Sizable
 effects of the hyperon-nucleon final state interaction are
found  near the $K^+\Lambda N$ and $K^+\Sigma N$ thresholds 
in the inclusive reaction.
Angular distributions for the exclusive process 
show clear $YN$ final state interaction effects in certain
kinematic regions.  Precise data especially for the inclusive
process around the
$K^+\Sigma N$ threshold would help to clarify the 
strength and property of the
$\Lambda N$-$\Sigma N$ interaction. 
\end{abstract}


\narrowtext
\abovedisplayskip 7mm
\belowdisplayskip 7mm
\abovedisplayshortskip 7mm
\belowdisplayshortskip 7mm
\jot 5mm
\newfont{\myfont}{cmti12 scaled \magstep1}

\section{Introduction}
\label{intro} 

Despite many investigations in the realm of hypernuclear physics
\cite{overview} many
properties of the hyperon-nucleon interaction remain  
uncertain. In the case of the $NN$ forces one has the rich set of $NN$
scattering data at one's disposal to adjust $NN$ force parameters.
Such a set is basically absent in the $YN$ system.
If one had had to extract the $NN$ force properties from spectra
of nuclei only, our knowledge on the $NN$ forces would have been remained
rather uncertain and limited. Few-body systems which can be solved
rigorously can  therefore play a
helpful role to acquire more detailed information on the $YN$ forces
including the important $\Lambda$-$\Sigma$ conversion.
A first example is the hypertriton, where recent rigorous solutions
\cite{ours}
of the coupled $\Lambda NN$-$\Sigma NN$ Schr\"odinger equation allowed to
exclude certain $YN$ forces which do not bind the hypertriton, assuming
that that 3-body $YNN$ forces
are absent.

Calculations coming up on $^4_\Lambda$He and $^4_\Lambda$H
\cite{hijama} will supplement these studies  on
$^3_\Lambda$H and, because of their richer spectra \cite{4b:exp},
will be even more informative. For the purpose of 
investigating 
$\Lambda$-$\Sigma$ conversion, scattering processes which cover a wide
range of energies from the $\Lambda$ to the $\Sigma$ threshold and
beyond appear to be especially informative. 

In a recent study \cite{tpole} the $S$-matrix  pole structure for the
$YN$ system has been investigated for various presently used $YN$ forces.
As is well known there is no bound state in the $\Lambda(\Sigma)N$
system,
but the present potential models support poles of the $S$-matrix
which are close to the $\Lambda$ and $\Sigma$ thresholds. Near the
$\Lambda$ threshold there are two $S$-wave virtual states 
at about $-3$ and $-5$ MeV,  and close
to the $\Sigma$ threshold there is a $^3{\rm S}_1\,$--$\,^3{\rm D}_1$
pole which appears at
different unphysical sheets of the Riemann energy surface depending
on the potential used.  This pole causes cusp-like structures in the
$\Lambda N$ scattering at the $\Sigma$ threshold. Their forms
and strengths depend on the potential employed.

Since performing hyperon-nucleon scattering experiments is very difficult, 
hyperon production processes on the deuteron, such as $\gamma( d,K^+)YN$, 
appear as natural candidates that allow exploring the the $YN$ interaction.
The hope is that the
pole structure of the $YN$ $t$-operator will have visible effects in such
a production process. 
Pioneering work in 
inclusive and exclusive $K^+$ photoproduction
on the deuteron has been
done before \cite{wright} based on simple hyperon-nucleon forces.
These calculations suggested that significant $YN$ final-state
interaction effects be present near the production thresholds.
Inclusive electron induced $K^+$ production on the deuteron 
using modern $YN$ forces appeared in \cite{lee}.
In this article we  reexamine 
the inclusive and exclusive  photoproduction processes
using various recently formulated  $YN$ forces \cite{nsc89,nsc97}
together 
with realistic  $NN$ forces and an  updated elementary photoproduction
operator \cite{bennhold} of the $K^+Y$ pair on a nucleon.
For the convenience of the reader 
the form of the production operator is presented in Sec.~II.
Section~III describes
the evaluation of the nuclear matrix element and the
inclusive cross section, while
Section~IV is devoted to the exclusive cross section.
Our numerical results for the various $YN$ forces are
displayed in Sec.~V. We conclude in Sec.~VI.

\section{The production operator}

Almost all analyses of kaon photoproduction on the nucleon
were performed at tree level \cite{saghai} in an effective Lagrangian
approach.
While this leads to violation of unitarity,
this kind of isobaric model 
provides a simple tool to parameterize kaon photoproduction 
off the nucleon because it is relatively easy to calculate and to use for 
production on nuclei. 
Without rescattering contributions the $T$-matrix is simply
approximated by the driving term alone which is assumed to be given by 
a series of tree-level diagrams.
The selected Feynman diagrams 
 for the $s$-, $u$-, and $t$-channel contain some unknown coupling
parameters to be adjusted in order to reproduce experimental data. 
Final state interaction is effectively absorbed in these
coupling constants which then cannot easily be compared to couplings
from other reactions.  
Guided by recent coupled-channel results\cite{feuster},
Ref.~\cite{elba} has reanalyzed the newest data\cite{saphir}
and constructed a tree-level amplitude that reproduces all available
$K^+ \Lambda$,
$K^+ \Sigma^0$ and $K^0 \Sigma^+$ photoproduction 
data and thus provides an effective parameterization
of these processes. The background terms included the standard
$s$-, $u$-, and $t$-channel contributions along with a contact
term that was required to restore gauge invariance after hadronic
form factors had been introduced\cite{haberzettl}.
 This model included the three nucleon resonances that have been found
in the coupled-channels approach to decay into the $K \Lambda$ channel,
the 
$S_{11}$(1650), $P_{11}$(1710), and $P_{13}(1720)$. For $K \Sigma$
production
further contributions from the $S_{31}$(1900) and $P_{31}$(1910)
 $\Delta$ 
resonances were added. 

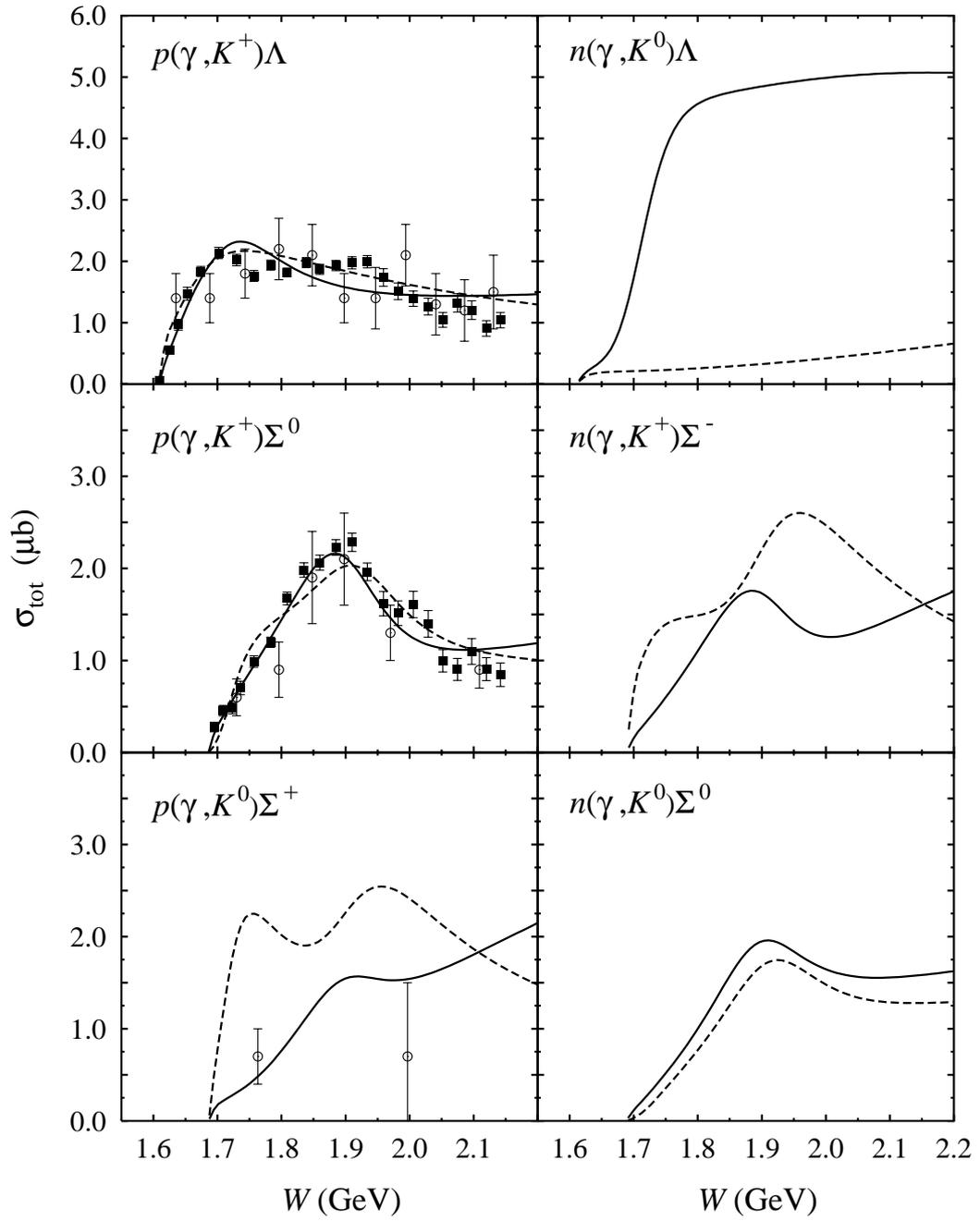
\begin{figure}
\caption{
The six possible kaon photoproduction channels compared
to new experimental data. The solid curve shows the model of
Ref.~\protect\cite{elba} while the dashed line shows an older calculation
of Ref.~\protect\cite{bennhold}. The new {\footnotesize SAPHIR} 
	data \protect\cite{saphir} are denoted by the solid 
      	squares, old data \protect\cite{old_data} are shown by the 
	open circles.}
\label{gamN}
\end{figure}

 Figure \ref{gamN} shows total cross section calculations for the
six different possible kaon photoproduction processes along with the 
newest data\cite{saphir}, 
comparing the model developed in Ref.~\cite{elba} 
with an older version\cite{bennhold}.
In order to obtain predictions for the
reactions on the neutron, isospin symmetry was used for the strong
vertices
while the electromagnetic resonance couplings of the neutron were taken
from the listing of helicity amplitudes in the Particle Data Tables. 
Below, we will only use the model of Ref.~\cite{elba} for the
calculations on the deuteron. 

The relativistic operator has the form
\begin{eqnarray}
t_{\gamma K}&=&\left(\frac{E_N+m_N}{2m_N}\right)^{\frac{1}{2}}
\left(\frac{E_Y+m_Y}{2m_Y}\right)^{\frac{1}{2}}\sqrt{\frac{m_Y}{E_Y}}
\sqrt{\frac{m_N}{E_N}}\times\nonumber\\
&&[\ 
   {\cal F}_1    \Vec{\sigma}\cdot\Vec{\epsilon}
+ {\cal F}_4    \Vec{\sigma}\cdot\Vec{p}_\gamma\,\Vec{p}_N\cdot
                                                           \Vec{\epsilon}
+ {\cal F}_5    \Vec{\sigma}\cdot\Vec{p}_\gamma\,\Vec{p}_Y\cdot
                                                           \Vec{\epsilon}
+ {\cal F}_8    \Vec{\sigma}\cdot\Vec{p}_N\,\Vec{p}_N\cdot\Vec{\epsilon}
+ {\cal F}_9    \Vec{\sigma}\cdot\Vec{p}_N\,\Vec{p}_Y\cdot\Vec{\epsilon}
\nonumber\\
&&
+ {\cal F}_{12} \Vec{\sigma}\cdot\Vec{p}_Y\,\Vec{p}_N\cdot\Vec{\epsilon}
+ {\cal F}_{13} \Vec{\sigma}\cdot\Vec{p}_Y\,\Vec{p}_Y\cdot\Vec{\epsilon}
+ {\cal F}_{14} \Vec{\sigma}\cdot\Vec{\epsilon}\,\Vec{\sigma}\cdot
                                \Vec{p}_\gamma\Vec{\sigma}\cdot\Vec{p}_N
+ {\cal F}_{15} \Vec{\sigma}\cdot\Vec{p}_Y\,\Vec{\sigma}\cdot
                                                         \Vec{\epsilon}
                \Vec{\sigma}\cdot\Vec{p}_\gamma \nonumber\\
&&
+ {\cal F}_{16} \Vec{\sigma}\cdot\Vec{p}_Y\,\Vec{\sigma}\cdot
                                                         \Vec{\epsilon}
                \Vec{\sigma}\cdot\Vec{p}_N
+ {\cal F}_{19} \Vec{\sigma}\cdot\Vec{p}_Y\,\Vec{\sigma}\cdot
                                                        \Vec{p}_\gamma
                \Vec{\sigma}\cdot\Vec{p}_N\,\Vec{p}_N\cdot\Vec{\epsilon}
\nonumber \\
&&
+ {\cal F}_{20} \Vec{\sigma}\cdot\Vec{p}_Y\,\Vec{\sigma}\cdot
                                                         \Vec{p}_\gamma
                \Vec{\sigma}\cdot\Vec{p}_N\,\Vec{p}_Y\cdot\Vec{\epsilon}
\ ] ~,
\label{op1} 
\end{eqnarray}
where $m_N$ and $m_Y$ denote the nucleon and hyperon masses, $E_N$ and
$E_Y$
their energies, $\Vec{p}_\gamma$, $\Vec{p}_N$ and $\Vec{p}_Y$ the photon,
nucleon and hyperon
momenta and $\Vec{\epsilon}$ the photon polarization. These momenta
together with the $K^+$ momentum  $\Vec{p}_K$ are
constrained by three-momentum conservation but otherwise the operator
$t_{\gamma K}$ is off-the-energy-shell when used inside a nucleus.
Since the elementary process is described using Feynman diagrams -- rather 
than multipole amplitudes -- the operator can be taken off-mass shell in a 
straightforward manner.
The amplitudes ${\cal F}_i$ are given in terms of kinematical quantities
and amplitudes  $A_i$  which are related to the various tree diagrams. 
The rather lengthy expressions  of ${\cal F}_i$ and  $A_i$
can be found in Ref. \cite{fa}. Note that 
the expression in Eq. (\ref{op1}) is valid in any frame.

It can be obviously rewritten into another form, which is also very
convenient for applications
\begin{eqnarray} 
t_{\gamma K} &=& i\left(\,L\,+\,i\Vec{\sigma}\cdot\Vec{K}\,\right) ~,
\label{op2} 
\end{eqnarray}
where 
\begin{eqnarray}
L&=& N \{
-( {\cal F}_{14} + {\cal F}_{15} - {\cal F}_{16} )\,
 \Vec{p}_N\cdot(\Vec{p}_\gamma\times\Vec{\epsilon})
+{\cal F}_{15}\,\Vec{p}_K\cdot(\Vec{p}_\gamma\times\Vec{\epsilon})
-{\cal F}_{16}\,\Vec{p}_N \cdot(\Vec{p}_K\times\Vec{\epsilon})
\nonumber \\
&&
-[({\cal F}_{19}+{\cal F}_{20})\,\Vec{p}_N\cdot\Vec{\epsilon} - 
{\cal F}_{20}
\Vec{p}_K \cdot \Vec{\epsilon}]\,\Vec{p}_N\cdot(\Vec{p}_K\times
\Vec{p}_\gamma)
\}
\end{eqnarray}
and
\begin{equation} 
\Vec{K}=-N ( T_1 \Vec{\epsilon} + T_2 \Vec{p}_\gamma + T_3 \Vec{p}_N 
+ T_4 
\Vec{p}_K ) ~,
\end{equation}
with
\begin{eqnarray}
T_1
&=& {\cal F}_{1}
+( {\cal F}_{14} - {\cal F}_{15} - {\cal F}_{16} )\,
\Vec{p}_N\cdot\Vec{p}_\gamma 
+{\cal F}_{15}\, (\Vec{p}_K\cdot\Vec{p}_\gamma -\Vec{p}_\gamma^2)
+{\cal F}_{16}\, (\Vec{p}_N\cdot\Vec{p}_K -\Vec{P}_N^2) ~,\\
%
T_2
&=& [ 
{\cal F}_{4}+{\cal F}_{5}+{\cal F}_{12}+{\cal F}_{13}-{\cal F}_{14}
+{\cal F}_{15}+{\cal F}_{16}
+(\Vec{p}_N\cdot\Vec{p}_K -\Vec{p}_N^2) ( {\cal F}_{19}+{\cal F}_{20} )
    \, ] \, \Vec{p}_N\cdot\Vec{\epsilon}
\nonumber\\
&&
-[
{\cal F}_{5}+{\cal F}_{13}+{\cal F}_{15}
+(\Vec{p}_N\cdot\Vec{p}_K -\Vec{p}_N^2){\cal F}_{20} 
\,]\, \Vec{p}_K\cdot\Vec{\epsilon} ~, \\
%
T_3
&=& [ 
{\cal F}_{8}+{\cal F}_{9}+{\cal F}_{12}+{\cal F}_{13}+2{\cal F}_{16}
+(2\Vec{p}_N\cdot\Vec{p}_\gamma +\Vec{p}_\gamma^2
 -\Vec{p}_K\cdot\Vec{p}_\gamma) ( {\cal F}_{19}+{\cal F}_{20} )
  \, ] \, \Vec{p}_N\cdot\Vec{\epsilon}
\nonumber\\
&&
-[
{\cal F}_{9}+{\cal F}_{13}+{\cal F}_{16}
+(2\Vec{p}_N\cdot\Vec{p}_\gamma +\Vec{p}_\gamma^2
  -\Vec{p}_K\cdot\Vec{p}_\gamma) {\cal F}_{20}
\,]\, \Vec{p}_K\cdot\Vec{\epsilon} ~,\\
T_4
&=& -[ 
{\cal F}_{12}+{\cal F}_{13}+{\cal F}_{16}
+\Vec{p}_N\cdot\Vec{p}_\gamma ( {\cal F}_{19}+{\cal F}_{20} )
    \, ] \, \Vec{p}_N\cdot\Vec{\epsilon}
+(
{\cal F}_{13}
+\Vec{p}_N\cdot\Vec{p}_\gamma {\cal F}_{20} 
  \, )\,  \Vec{p}_K\cdot\Vec{\epsilon} ~.
\end{eqnarray}

Furthermore, we have 
\begin{equation} 
N= 
\left(\frac{E_N+m_N}{2m_N}\right)^{\frac{1}{2}}
\left(\frac{E_Y+m_Y}{2m_Y}\right)^{\frac{1}{2}}
\sqrt{\frac{m_Y}{E_Y}}
\sqrt{\frac{m_N}{E_N}} ~.
\end{equation}

%

\section{The inclusive cross section}
\label{inclusive}
The cross section for the inclusive process $d( \gamma , K^+)$ is given as
\begin{eqnarray}
d\sigma&=&
\frac{1}{6} \sum_{Y} 
\sum_{\mu_d\,\epsilon}
\sum_{\mu_Y\,\mu_N}
 \sum_{\nu_Y\, \nu_N}
 \frac{(2\pi)^3}{4E_K E_\gamma}
\int\frac{d\Vec{p}_K}{(2\pi)^3}\frac{d\Vec{p}_Y}{(2\pi)^3}
\frac{d\Vec{p}_N}{(2\pi)^3} \nonumber \\
&&\times \left|\,\sqrt{2} \langle \,\Psi^{(-)}_{\Vec{q}_Y \mu_Y\, \nu_Y\, 
\mu_N\, \nu_N}
\,|\,t_{\gamma K}(1)
\,|\,\Psi_d\, \mu_d\,\rangle \,\right|^2 \times (2\pi)^4 \delta^{(4)}
( P_d + Q - p_Y - p_N ) ~,
\label{ic1}
\end{eqnarray}
where the $\mu$'s and $\nu$'s are spin and isospin  magnetic quantum
 numbers and $\epsilon$ denotes the
two photon polarizations. The states $\Psi$ refer to the two baryons
only and $\Vec{q}_Y$ is the (nonrelativistic) relative momentum of the
final hyperon and nucleon. The sum over $Y$ refers to the $\Lambda$ and
$\Sigma$ channels. The dependencies on the $K^+$ and photon parameters,
aside from the normalization factors shown explicitly,
are absorbed into the $t_{\gamma K}$ operator. 
In Eq. (\ref{ic1}) we have also introduced
the momentum transfer $Q = p_\gamma - p_K$. The factor $\sqrt{2}$ comes
from proper antisymmetrization and the argument 
1 in $t_{\gamma K}(1)$ indicates
that it acts only on particle 1, which in the final state is given by
the hyperon. Equation (\ref{ic1}) can be easily derived  via Feynman rules
but using (inconsistently) nonrelativistic two-baryon wavefunctions.
This derivation has the nice feature that it shows how to use the single 
particle operator introduced in Sec.~II between the wavefunctions.
Note that the factors $\displaystyle \sqrt{\frac{m_Y}{E_Y}}
\sqrt{ \frac{m_N}{E_N}}$ in Eq.~(\ref{op1})
should be kept as a part of the operator.
We point out that we have used plane waves for the kaons.
While the final-state interaction of the kaon with the hyperon
is effectively absorbed in the elementary amplitude the interaction
of between the kaon and the spectator nucleon is neglected. However,
since the $K^+ N$ interaction is rather weak on a hadronic scale we expect
the effect of this omission to be negligible.
In the nuclear matrix element appearing in Eq.~(\ref{ic1})
the corresponding hyperon and nucleon momenta are integrated over.

The kinematics and the elementary operator is kept in its relativistic
form.  Though this is somewhat inconsistent in relation to the
nonrelativistic
wave functions we believe it is a step in the right direction. 
The estimates\cite{wright}  for relativistic effects related to the
deuteron 
wavefunction
turned out to be insignificant.

We note that we shall work throughout in the zero total momentum frame of
the final two baryons.
The integrations in Eq. (\ref{ic1}) can be easily carried out
in the c.m. frame of the final two baryons.
To do that we supplement the expression~(\ref{ic1})
with 
$\displaystyle \frac{m_Y}{E_Y}\frac{m_N}{E_N}
  \times \frac{E_Y}{m_Y}\frac {E_N}{m_N}$.
The second factor multiplies the nuclear matrix
element, which again is treated non-relativistically.
Thus, the second factor is replaced by unity. We
end up with the result
\begin{eqnarray}
\frac{d\sigma}{dp_K d\Omega_K}&=&
\frac{\Vec{p}^2_K}{(2\pi)^2\, 4E_\gamma E_K W } \sum_Y
 m_Y m_N \, |\Vec{q}_Y| \nonumber \\
&&\times \frac{1}{6}
 \sum_{\mu_d\, \epsilon}
 \sum_{\mu_Y\,\mu_N}
\sum_{ \nu_Y\,\nu_N}
\int d\hat{\Vec{q}}_Y \left|\,\sqrt{2}\,
\langle \,\Psi^{(-)}_{\Vec{q}_Y 
\mu_Y\nu_Y\mu_N\nu_N}\,|\,t_{\gamma K}(1)\,|\,
\Psi_d \mu_d\,\rangle \,\right|^2 ~,
\label{ic2}
\end{eqnarray}
where $W^2=(P_d+Q)^2$ and $|\Vec{q}_Y|$ is determined by the energy
conserving
delta function.
The nuclear matrix element can be conveniently rewritten  by applying
the M\"oller wave operator generating the final scattering state
 to the right.  One obtains 
\begin{eqnarray}
\langle \,\Psi^{(-)}_{\Vec{q}_Y
\mu_Y\nu_Y\mu_N\nu_N}\,|\,t_{\gamma K}(1)\,|\,\Phi_d \mu_d\,\rangle &=&
\langle \,\Vec{q}_Y \mu_Y\nu_Y\mu_N\nu_N
\,|\,(1+tG_0)t_{\gamma K}(1)\,|\,\Psi_d \mu_d\,\rangle  \nonumber \\
&\equiv&\langle \,\Vec{q}_Y
\mu_Y\nu_Y\mu_N\nu_N\,|\,T\,|\,\Psi_d \mu_d\,\rangle ~.
\label{ic3}
\end{eqnarray}
Since we allow for $\Lambda$-$\Sigma$ conversion, the state
$\langle \Psi^{(-)}_{\Vec{q}_Y\mu_Y\nu_Y\mu_N\nu_N}\,|$
is a row with a $\Lambda$ and a $\Sigma$ component.
Similarly the free state 
$\langle \Vec{q}_Y \mu_Y\nu_Y\mu_N\nu_N\,|$
has two components
($\langle \Vec{q}_{\Lambda} \mu_{\Lambda}\nu_{\Lambda}\mu_N\nu_N\,|\,
 ,\, 0 )$ for $Y=\Lambda$ and
($0\, ,\, \langle \Vec{q}_{\Sigma} \mu_{\Sigma}\nu_{\Sigma}\mu_N\nu_N\,|
\, )$
 for $Y=\Sigma$.
The operators $t$ and $G_0$ occurring in Eq.~(\ref{ic3}) are $2\times 2$
matrices acting on $\Lambda$ and $\Sigma$ components.
The operator
$t_{\gamma K}(1)$
converts a nucleon into a hyperon and is therefore like the operator 
$T$ a two-component object in $\Lambda$ and $\Sigma$ space. 

Obviously, $T$ applied to the deuteron state
obeys the integral equation
\begin{equation}
T\,|\,\Psi_d \mu_d\,\rangle \,=\,t_{\gamma K}(1)\,|\,\Psi_d 
\mu_d\,\rangle 
\,+\,V G_0 T \,|\,\Psi_d \mu_d\,\rangle ~,
\label{ic4}
\end{equation}
where $V$ is the hyperon-nucleon force including $\Lambda$-$\Sigma$
conversion and $t$ in Eq.~(\ref{ic3}) is the corresponding $t$-operator
as given by a Lippman-Schwinger equation. 
The energy entering
the free two-baryon propagator $G_0$ which is a diagonal matrix
is given non-relativistically  as
\[
e_Y = (P_d+Q)^2 - m_N-m_Y ~.
\] 
Let us now define
\begin{equation}
T\,|\,\Psi_d \mu_d\,\rangle \,=\,\pmatrix{T_\Lambda|\,\Psi_d 
\mu_d\,\rangle \cr
        T_\Sigma|\,\Psi_d \mu_d\,\rangle \cr} ~.
\label{ic5}
\end{equation}
Then Eq.~(\ref{ic4}) takes the form of a coupled set:
\begin{equation}
T_Y\,|\,\Psi_d \mu_d\,\rangle \,=\,t_{\gamma K}^Y (1)\,|\,\Psi_d 
\mu_d\,\rangle 
\,+\,
\sum_{Y^\prime}
V_{Y,Y'} G_0^{Y'} T_{Y'} \,|\,\Psi_d \mu_d\,\rangle ~.
\label{ic6}
\end{equation}
Equation~(\ref{ic6}) contains the elementary operator
$t^Y_{\gamma K}$
producing a specific hyperon $Y$.
We solve this set, Eq.~(\ref{ic6}), in momentum space and partial
 wave decomposed. 
We introduce the basis $\langle q_Y \alpha|= 
\langle q_Y ( l s ) j m t m_t|$
 with $l,s,j(m)$ being the relative orbital, total spin and total
 angular momentum (with magnetic quantum number) of the two baryon 
system and $t (m_t)$ the total two-baryon isospin (with magnetic
 quantum number).
Then  Eq.~(\ref{ic6}) turns into
\begin{eqnarray}
\langle \,q_Y\,\alpha\,|\,T_Y\,|\,\Psi_d \mu_d\,\rangle 
&=&\langle \,q_Y\,\alpha\,|\,t_{\gamma K}^Y (1)\,|\,\Psi_d 
\mu_d\,\rangle  
\nonumber \\
&&+\sum_{Y^\prime}\sum_{\alpha^\prime}\int^\infty_0 dq^\prime_{Y'}
 q^{\prime 2}_{Y'}
\langle \,q_Y\,\alpha\,|\,V_{YY'}\,|\,q^\prime_{Y'},\alpha^\prime
\,\rangle  
\nonumber \\
&&\hspace{10mm}\times \frac{1}{e_Y-\frac{q^{\prime 2}_{Y'}}
{2 \mu_{Y^{\prime}}}+i\epsilon}\,
\langle \,q^\prime_{Y'}\,\alpha^\prime\,|\,T_{Y'}\,|\,\Psi_d \mu_d
\,\rangle ~.
\label{ic7}
\end{eqnarray}
Note that the operator
$t_{\gamma K}^Y (1)$
includes hyperon production on the proton and the neutron.
Consequently, the resulting two baryons can be of different types.
Applying the operator 
$t_{\gamma K}$ on $|\,\Psi_d\rangle $ with the isospin part of the
deuteron written out explicitly yields
\begin{eqnarray}
\lefteqn{t_{\gamma K}(1)\,|\,\Psi_d \mu_d\,\rangle } \nonumber \\
&=&t_{\gamma k}(1){1\over\sqrt{2}}
\left(|\,p(1)\rangle |\,n(2)\rangle -|\,n(1)\rangle |\,p(2)\rangle
 \right)|\,\Phi_d \mu_d\,\rangle 
\nonumber \\
&=&{1\over\sqrt{2}}\Biggl(
|\,\Lambda(1)\rangle |\,n(2)\rangle \langle \Lambda\,|t_{\gamma k}|\,p
\rangle 
\, +\, |\,\Sigma^0(1)\rangle |\,n(2)\rangle \langle 
\Sigma^0\,|t_{\gamma k}|\,p\rangle \nonumber \\
&&\quad\quad\quad
-|\,\Sigma^{-}(1)\rangle |\,p(2)\rangle \langle \Sigma^{-}
\,|t_{\gamma k}|\,n\rangle \Biggr)
|\,\Phi_d \mu_d\,\rangle ~.
\label{ic8}
\end{eqnarray}

The main task is the evaluation of the driving term  in Eq.~(\ref{ic7})
based on the elementary operator $t_{\gamma K}$ from Sec.~II.
 This single particle operator acts in the two-baryon space.
 We introduce the relative momentum between general hyperon and nucleon
momenta $\Vec{k}_Y$ and $\Vec{k}_N$ as
\begin{equation} 
\Vec{q}_Y=\frac{m_N \Vec{k}_Y - m_Y \Vec{k}_N}{m_N+m_Y} ~,
\label{ic9}
\end{equation} 
and the relative momentum between general two nucleon momenta
$\Vec{k}_1$ and 
$\Vec{k}_2$
as
\begin{equation} 
\Vec{q}=\frac{1}{2}(\Vec{k}_1-\Vec{k}_2) ~.
\label{ic10}
\end{equation} 
Then we obtain in obvious notation
\begin{eqnarray}
\langle \,q_Y\,\alpha\,|\,t_{\gamma K}^Y (1)\,|\,\Psi_d \mu_d\,\rangle 
&=& \int d\hat{\Vec{q}}_Y \left[\,Y_l^*(\hat{\Vec{q}}_Y)
\otimes\, \langle \,s\,| \,\right]^j_m
\, \sum_{\tilde Y} \, C^{\, tY\tilde Y}\,\, t_{\gamma K}^{\tilde Y}
(\Vec{k}_Y,\Vec{k}_1) 
 \nonumber
 \\
&& \hspace{10mm} \times
\sum_{l_d=0,2} \left[\,Y_{l_d}(\hat{\Vec{q}}) 
\otimes \,|\,s_d\,\rangle \,\right]^{j_d}_{\mu_d}
\phi_{l_d}(q) ~,
\label{ic11}
\end{eqnarray}
where the sum over $\tilde Y$ refers to the $\Lambda$, $\Sigma^0$ and 
$\Sigma^-$
production processes and according to Eq.~(\ref{ic8}) the coefficient
$C^{\, tY\tilde Y}$ is given as 
\begin{eqnarray}
 C^{\, tY\tilde Y} &=& \left\{
\begin{array}{rcl}
        {1\over\sqrt{2}}C(0\,{{\textstyle \frac{1}{2}}}\,t,0\,
	-{{\textstyle \frac{1}{2}}}\,m_t) & ~~~~~ &
        \textrm{for $Y=\Lambda$ and $\tilde Y=\Lambda$} \\ \\
        {1\over\sqrt{2}}C(1\,{{\textstyle \frac{1}{2}}}\,t,0\,
	-{{\textstyle \frac{1}{2}}}\,m_t) &  &
        \textrm{for $Y=\Sigma$ and $\tilde Y=\Sigma^0$} ~~~,\\ \\
        -{1\over\sqrt{2}}C(1\,{{\textstyle \frac{1}{2}}}\,t,-1\,
	{{\textstyle \frac{1}{2}}}\,m_t) & &
        \textrm{for $Y=\Sigma$ and $\tilde Y=\Sigma^-$}
               \end{array}
\right.
\label{ic12}
\end{eqnarray}
with $m_t=-{1\over2}$.

 For the sake of transparency we kept the notation for the 
 momenta occurring on the right-hand side of Eq.~(\ref{ic11})
with their obvious meanings. But they should be expressed in terms
of the relative momentum $\Vec{q}_Y$ and a given external
momentum  $\Vec{Q}$ as
%
\begin{eqnarray}
\label{ic13} \Vec{k}_Y &=& \Vec{q}_Y ~,\\
\label{ic14} \Vec{k}_1 &=& \Vec{q}_Y-\Vec{Q} ~, \\
\label{ic20} \Vec{q} &=& \Vec{q}_Y - \frac{1}{2}\Vec{Q} ~.
\end{eqnarray}
%
Now the form of Eq.~(\ref{op2}) for $t_{\gamma K}$ is very
convenient and we obtain
\begin{eqnarray}
\langle \,q_Y\,\alpha\,|\,t_{\gamma K}^Y (1)\,|\,\Psi_d \mu_d\,\rangle 
&=&\sum_{\tilde Y} C^{tY\tilde Y}
\sum_{l_d}\sum_{m^\prime m^{\prime\prime}}
\,C(\,lsj,m^\prime\, m-m^\prime\,  m)
C(\,l_d s_d j_d, m^{\prime\prime}\, \mu_d -m^{\prime\prime}\, \mu_d)
\nonumber \\
&&\times \{\,
\delta_{ss_d} \delta_{m-m^\prime,\,\mu_d-m^{\prime\prime}} 
\int d\hat{\Vec{q}}_Y
Y_{lm^\prime}^*(\hat{\Vec{q}}_Y)\,i L^{\tilde Y}\,Y_{l_d m''}
(\hat{\Vec{q}})
\phi_{l_d}(q)  \nonumber \\
&&
-
\,\sum_{\nu} C(\,1 s_d s, \nu,\mu_d-m^{\prime\prime},m-m^{\prime}\,)\,
\frac{(-)^{s-s_d+1}}{\sqrt{\hat s}}
 \langle \,s\,||\,\sigma(1)\,||\,s_d\,\rangle  \nonumber \\
&&\hspace{10mm} \times \int d\hat{\Vec{q}}_Y
 Y_{lm^\prime}^*(\hat{\Vec{q}}_Y)\,(-)^\nu K^{\tilde Y}_{-\nu}
\,Y_{l_d m''}(\hat{\Vec{q}}) \phi_{l_d}(q)
\,\} ~,
\label{ic15}
\end{eqnarray}
where the index $\tilde Y$ for the operators
$L$ and $K_{-\nu}$
specifies the three individual production processes 
as already mentioned.
We have chosen  the reduced spin matrix element to have the form
\begin{equation}
\langle \,s||\sigma(1)||s_d\,\rangle =
\sqrt{\hat{s} \cdot 6 \cdot \hat{s}_d}\,(-)^{-s_d}
\,\left\{
\begin{array}{ccc}
s&s_d&1 \\
\frac{1}{2}&\frac{1}{2}&\frac{1}{2}\\
\end{array}
\right\} ~.
\label{ic16}
\end{equation}
The remaining two-fold integrals will be performed numerically.

Once the coupled set, Eq.~(\ref{ic7}), has been solved we obtain the
 hadronic matrix element from Eq.~(\ref{ic3}) as
\begin{eqnarray}
&&\langle \,\Psi^{(-)}_{\Vec{q}_Y \mu_Y \nu_Y \mu_N \nu_N}\,|
\,t_{\gamma K}(1)\,|\,\Psi_d \mu_d\,\rangle  
 \hfill \nonumber \\ 
&=& \sum_{lsjm t}\,\langle \,\Vec{q}_Y \mu_Y \nu_Y \mu_N 
\nu_N\,|\,(ls)jmtm_t\,\rangle 
\,\langle \, (ls)jmtm_t\,|\,T\,|\,\Psi_d \mu_d\,\rangle 
 \nonumber \\
&=& \sum_{lsjmt}\,C(lsj,m-\mu_Y-\mu_N,\mu_Y+\mu_N)
\, C({\textstyle \frac{1}{2}\frac{1}{2}}s,\mu_Y \mu_N)
 \nonumber \\
&&
\times Y_{l\,m-\mu_Y-\mu_N}(\hat{\Vec{q}}_Y)
\, C(t_Y {\textstyle \frac{1}{2}} t, \nu_Y\nu_N m_t)
\, \langle \,q_Y\,(ls)jmtm_t\,|\,T_Y\,|\,\Psi_d \mu_d\,\rangle ~.
\label{ic17}
\end{eqnarray}
The isospins $t_Y$  of the hyperon are 0 and 1
for $\Lambda$ and $\Sigma$, respectively.
This leads  to the final expression for the
inclusive cross section from Eq.~(\ref{ic2}) 
\begin{eqnarray}
\frac{d\sigma}{dp_K d\Omega_K}&=&
\frac{\Vec{p}^2_K}{(2\pi)^2\, 4E_\gamma E_K W } \sum_Y
 m_Y m_N \, |\Vec{q}_Y| \nonumber \\
&&\times \frac{1}{6}
 \sum_{\mu_d\, \epsilon}
 \sum_{lsjm}
\sum_{ t}
\left|\,\sqrt{2}\,\langle \,q_Y\,(ls)jmtm_t\,|\,T_Y\,|\,\Psi_d 
\mu_d\,\rangle \,
\right|^2 ~~.
\label{ic18}
\end{eqnarray}

It turns out that the convergence in $j$ is rather slow due to the
plane wave part of the amplitude.
Therefore, we treated that part separately without partial wave
 decomposition. Let $j_{\rm max}$ be the total two-baryon angular momentum
 beyond which the final state interaction can be neglected. Then
\begin{eqnarray}
&&\sum\limits_{l,s,\, j{\scriptscriptstyle >}
j_{\mbox{\tiny max},m}}
\left|\sqrt{2}\langle q_Y(ls)jmtm_t|T_Y|\,\Psi_d\mu_d\rangle 
\right|^2\nonumber\\
&=&
\sum\limits_{l,s,\, j{\scriptscriptstyle > }j_{\mbox{\tiny max},m}}
\left|\sqrt{2}\langle q_Y(ls)jmtm_t|t^Y_{\gamma
k}(1)|\,\Psi_d\mu_d\rangle \right|^2\nonumber\\
&=&\sum\limits_{\mu_Y\mu_N}\int d\hat{\Vec q}_Y \left|
\sqrt{2}\langle \mbox{\boldmath $q$}_Y\mu_Y\mu_Ntm_t|t_{\gamma
k}(1)|\,\Psi_d\mu_d\rangle \right|^2\nonumber\\
&&-\sum\limits_{l,s,\, j{\scriptscriptstyle\le}
j_{\mbox{\tiny max}},m}\left|\sqrt{2}
\langle q_Y(ls)jmtm_t\,|t^Y_{\gamma k}(1)|\Psi_d\mu_d\rangle \right|^2 ~~.
\label{ic19}
\end{eqnarray}
The second part is then added to the corresponding sum ($j\le j_{\rm max}$)
with $t_{\gamma K}^Y (1)$ replaced by $T_Y$. In this manner the
explicitly
partial wave projected part is due only to FSI.

\section{The exclusive cross section}
\label{exclusive}
The exclusive cross section for the process
 $\gamma( d,K^+ Y)N$
 follows easily from Eq.~(\ref{ic2}) as
\begin{eqnarray}
{d^5\sigma\over dp_k\,d\Omega_k\,d\Omega_Y}
&=&
\frac{\Vec{p}^2_K}{(2\pi)^2\, 4E_\gamma E_K} 
\frac{m_Y m_N \, |\Vec{q}_Y|}{W} 
  \nonumber \\
&&\times \frac{1}{6}
 \sum_{\mu_d\, \epsilon}
 \sum_{\mu_Y\,\mu_N}
\sum_{ \nu_Y\,\nu_N}
\left|\,\sqrt{2}\,
\langle \,\Psi^{(-)}_{\Vec{q}_Y 
\mu_Y\nu_Y\mu_N\nu_N}\,|\,t_{\gamma K}(1)\,|\,
\Psi_d \mu_d\,\rangle \,\right|^2 ~~.
\end{eqnarray}
The events have to lie on a kinematical locus, which relates the $K^+$
and $Y$ energies.
This relation
 is given by 
\begin{equation}
\left(\sqrt{\Vec{q}_Y^2+m_Y^2}+\sqrt{\Vec{q}_Y^2+m_N^2}\right)^2
=(P_d+Q)^2 ~~.
\end{equation}

Again we use a separation of the total amplitude into the plane wave part
which is treated without partial wave decomposition
and the part due to FSI which is partial wave decomposed. 
We write
\begin{eqnarray}
&&\sqrt{2}\langle \mbox{\boldmath$q$}_Y\mu_Y\nu_Y\mu_N\nu_N\,|T|
\,\Psi_d\mu_d\rangle 
\nonumber\\
&=&\sum\limits_s\sqrt{2}\,\,\, C({{\textstyle \frac{1}{2}}}{{\textstyle 
\frac{1}{2}}}s,\mu_Y\mu_Nm_s)
\langle sm_s|\langle \mbox{\boldmath$q$}_Y\nu_Y\nu_N|t_{\gamma k}(1)|
\,\Psi_d\mu_d\rangle 
\nonumber\\
&+&\sum\limits_{l,s,j,m} C({{\textstyle \frac{1}{2}}}{{\textstyle 
\frac{1}{2}}}s,\mu_Y\mu_N)
(lsj,m-\mu_Y-\mu_N,\mu_Y+\mu_N)Y_{lm-\mu_Y-\mu_N}(\hat{\Vec{q}}_Y)
\nonumber\\
&&
\quad\quad\quad\times
\, 
\sqrt{2}\langle q_Y(ls)jm\,\nu_Y\nu_N\,|
T_Y-t^Y_{\gamma k}(1)|\,\Psi_d\mu_d\rangle ~.
\end{eqnarray}

\section{Results}
\label{results} 

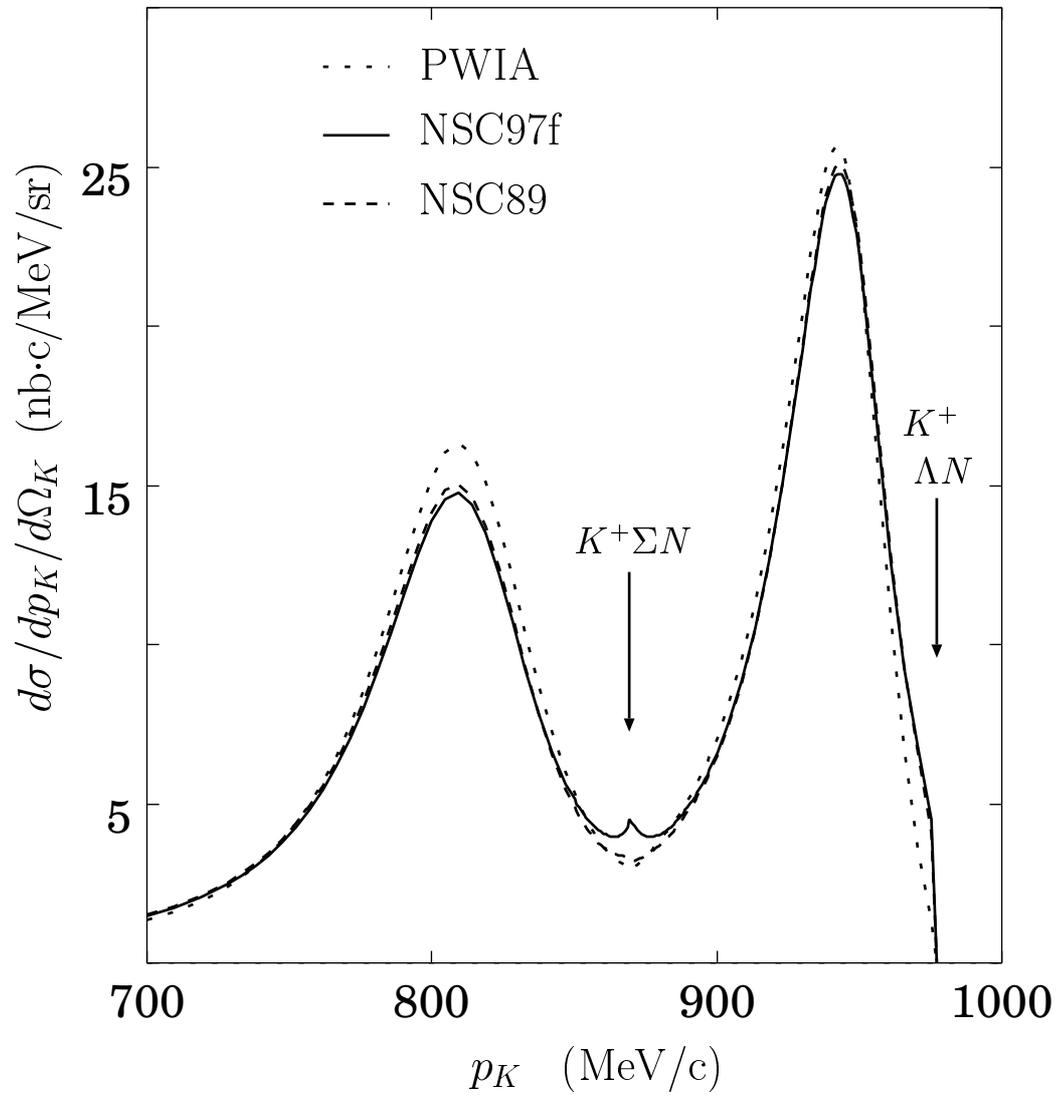
\begin{figure}
\caption{
The inclusive $\gamma( d,K^+)$ cross section as a function
of lab momenta $p_K$ for $\theta_K=0^\circ$ and 
photon lab energy $E_\gamma=1.3$ GeV.
 The plane wave result is compared to two $YN$ force predictions. 
The FSI effects are especially pronounced near the $K^+\Lambda N$ 
and $K^+\Sigma N$ thresholds the locations of which are indicated
by the arrows.
\label{fig1}}
\end{figure}

In this study we compare results
for the hyperon-nucleon forces NSC89\cite{nsc89} and 
NSC97f\cite{nsc97}. Both lead to the correct hypertriton binding
energy\cite{ours}. One of us (K.M.) recently investigated  various
new versions of the $YN$ forces developed by the Nijmegen
 group\cite{nsc97}. Only the NSC97f force binds the hypertriton
 correctly. 
The deuteron wavefunction is generated by the Nijmegen93 potential
\cite{nij93}.
The value $j_{\rm max}$ up to which FSI had to be taken into account
turned out to be $j_{\rm max}=2$. The results presented below are given at a 
photon energy $E_\gamma =1.3$ GeV. In Fig.~\ref{fig1}, we compare the
inclusive
cross sections for $d (\gamma , K^+)$ in plane wave impulse
approximation (PWIA) with calculations that include FSI.
In order to obtain the largest cross section
we have chosen $\theta _K=0^\circ$.
The two pronounced peaks around $p_K=$ 945 and 809 MeV/c can be
understood
in PWIA. They are due to quasi-free processes, where one of the 
nucleons in the deuteron is a spectator and has zero momentum in the 
lab system. This then leads to a vanishing argument $q=0$ in the
deuteron wavefunction, which causes the peaks. Under this condition
the kinematics of the $\gamma$-induced process on a single nucleon 
fixes the peak positions for $p_K$ in the lab system.
  
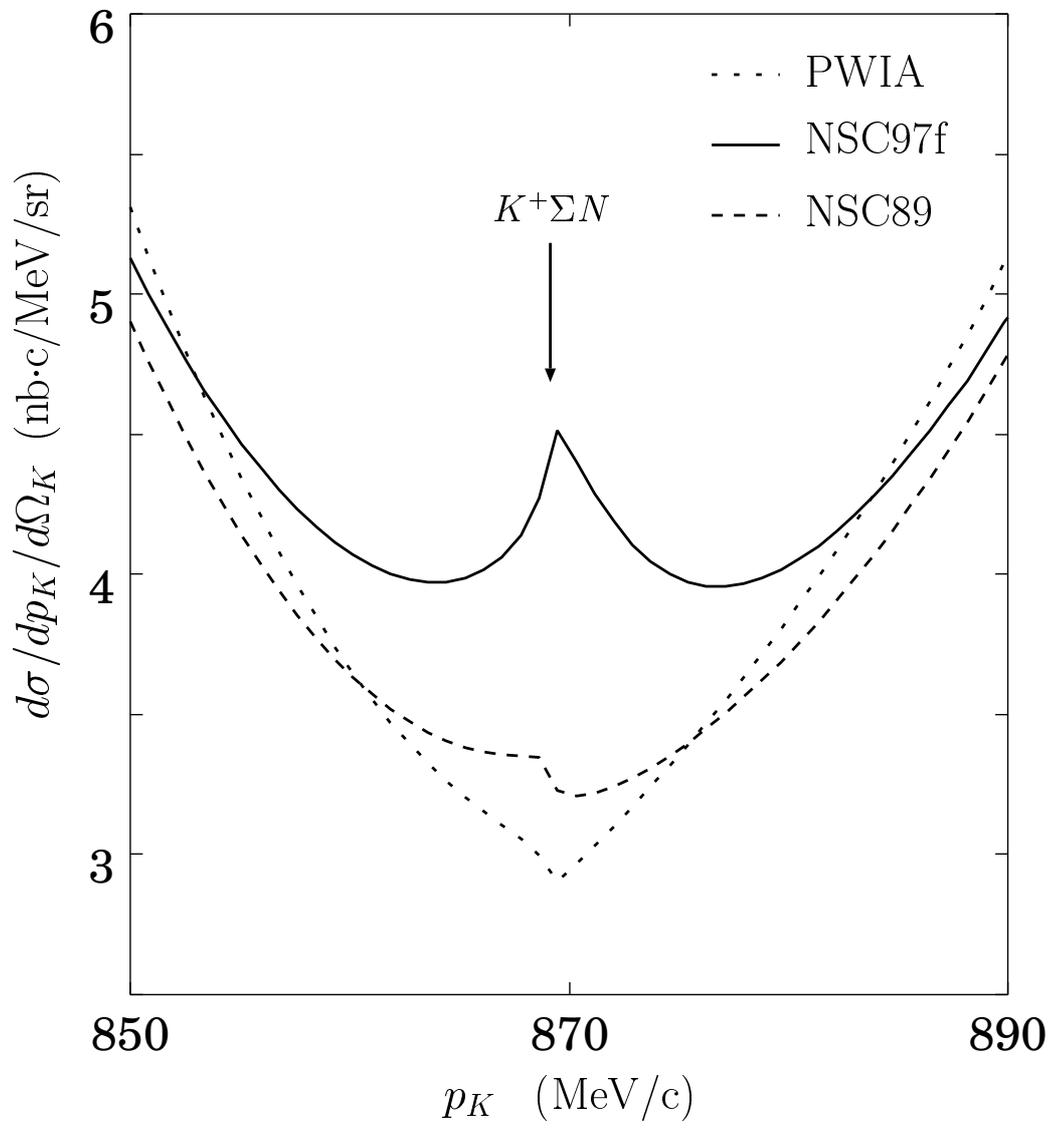
\begin{figure}
\caption{
The results of Fig.~\ref{fig1} enlarged around the $K^+\Sigma N$
threshold.
 \label{fig2}}
\end{figure}

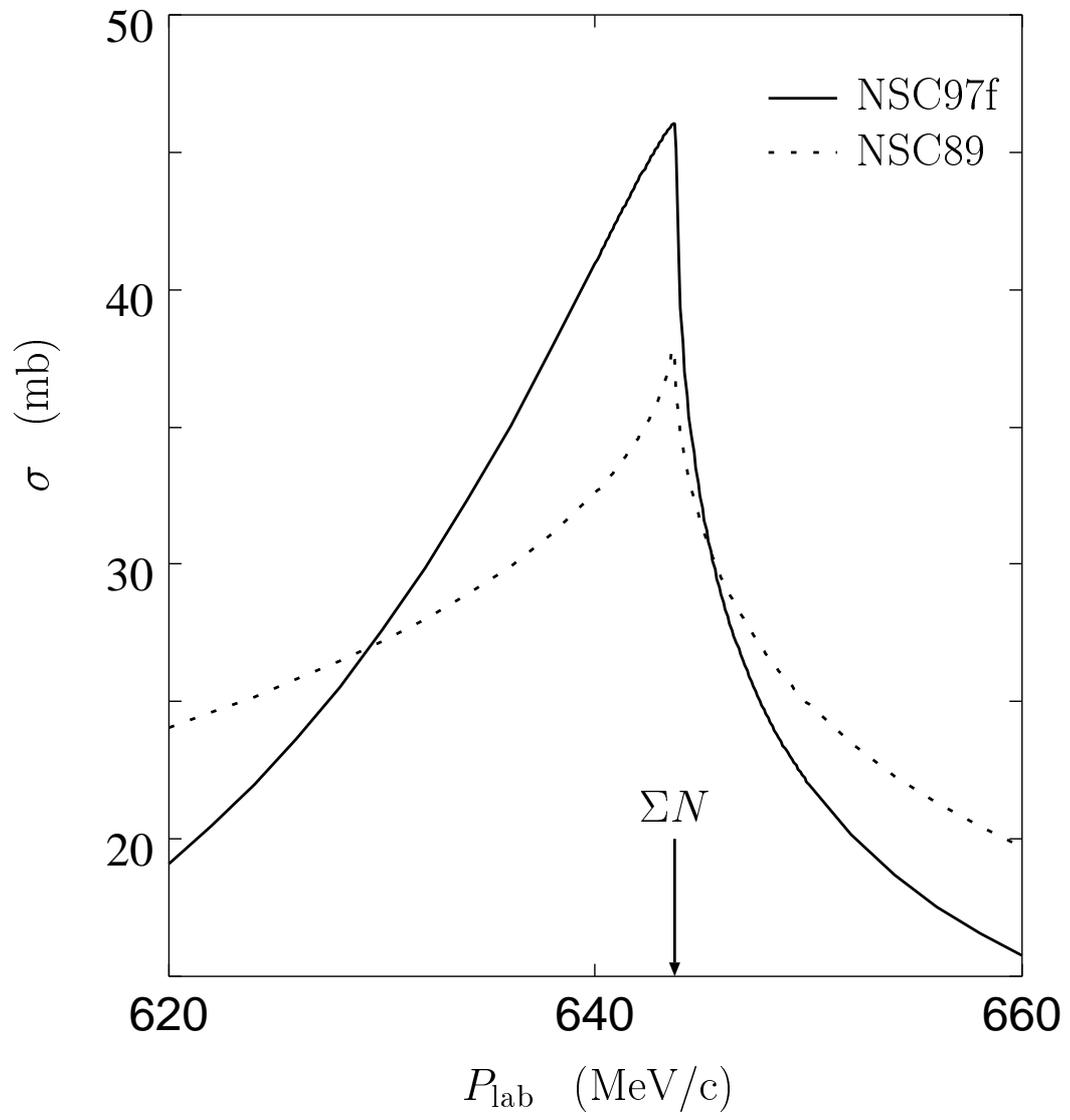
\begin{figure}
\caption{
The total $\Lambda N$ elastic cross section as a function of the
$\Lambda$ lab momentum around the $\Sigma N$ threshold indicated
by the arrow. The NSC97f
prediction leads to a more pronounced peak  structure than the NSC89
prediction.
 \label{fig3}}
\end{figure}

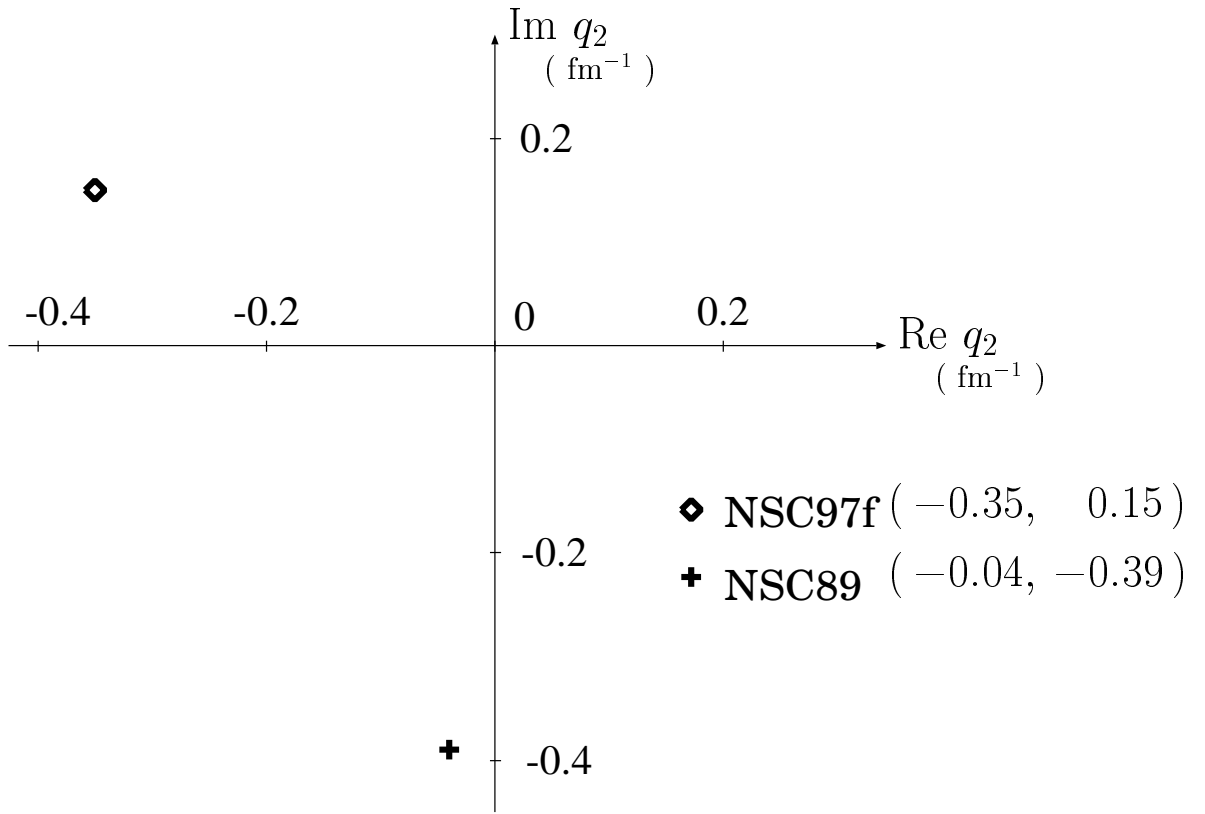
\begin{figure}
\caption{
The $S$-matrix pole positions for the $\Lambda N$-$\Sigma N$ system
in the complex $p_{\Sigma N}$ plane for the two modern $YN$ forces
NSC97f and NSC89 (see text).
\label{fig4}}
\end{figure}

We see deviations 
between the plane wave result and the results with
FSI based on the NSC89 and NSC97f hyperon-nucleon forces. 
Near the $K^+\Lambda N$ threshold FSI enhance the cross section
by up to 86\%. Near the $K^+\Sigma N$ threshold
the effects are also of interest.  
While NSC89 has hardly any effect,
NSC97f leads to a prominent cusp-like structure.
The neighborhood of the $K^+\Sigma N$ threshold
is shown again enlarged in Fig.~\ref{fig2}. The two $YN$ potentials
lead to 
predictions
which differ by up to 35\%. Different predictions of the two
potentials are also seen in the total elastic $\Lambda N$ cross section
as depicted in Fig.~\ref{fig3}. The peak for NSC97f is significantly
higher
 near the $\Sigma N$ threshold than for NSC89.
As worked out in Ref.~\cite{tpole}, this can be traced back to the
location
of the $S$-matrix pole for the $\Lambda N$-$\Sigma N$ system around the 
$\Sigma N$ threshold. We show in Fig.~\ref{fig4} the complex plane of the
relative
$\Sigma N$ momentum $p_{\Sigma N}$.  Each of the two $YN$ potentials
generates 
a pole in the state $^3{\rm S}_1\,$--$\,^3{\rm D}_1$ near
 $p_{\Sigma N}=0$.
The potential NSC89 leads to a pole position which in a single
channel
case would be called a virtual state (in this case it would lie
exactly on the imaginary axis).
 The coupling of the $\Lambda$ and $\Sigma$
channels moves the pole for the NSC97f force away from the positive
 imaginary axis into the second $p_{\Sigma N}$ quadrant. 
In a time-dependent description the energy related to that pole position 
leads to a decreasing amplitude. In the literature, this sort of 
pole is sometimes referred to as an `unstable bound state'. 
Apparently, the actual pole position depends on the details
of the $YN$ force.
The pole positions are an inherent property of the $YN$ forces and
the actual location chosen
by nature should be determined with the help of experimental measurements.

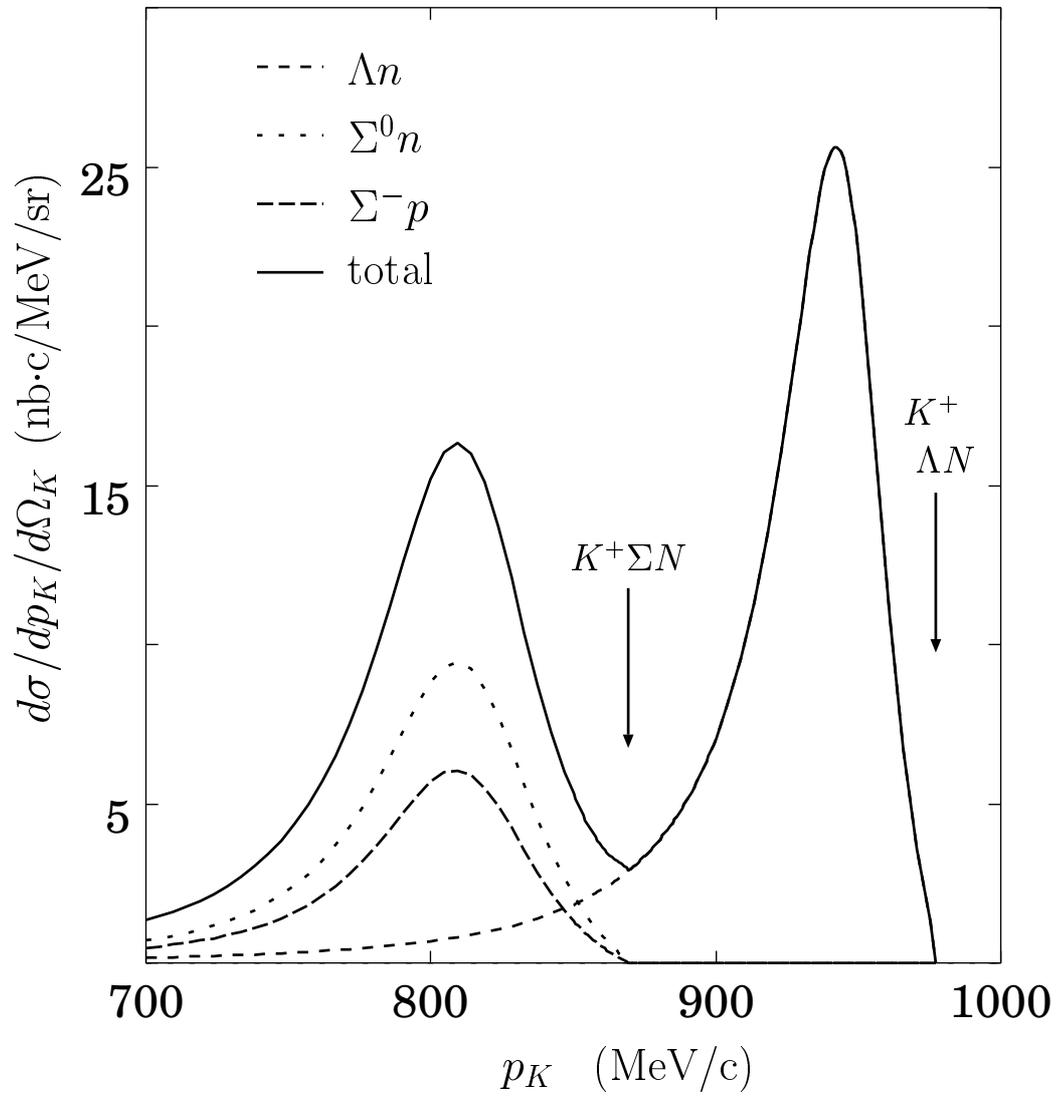
\begin{figure}
\caption{
The inclusive $\gamma( d,K^+)$ cross section as a function
of lab momenta $p_K$ for $\theta_K=0^\circ$ and 
photon lab energy $E_\gamma=1.3$ GeV in plane
wave approximation. 
The $K^+\Lambda N$ 
and $K^+\Sigma N$ thresholds are indicated
by the arrows.
The additive contributions for the $\Lambda n$,
$\Sigma^0 n$ and $\Sigma^- p$ processes are shown separately
 and summed up.
\label{fig5}}
\end{figure}

Another interesting insight into the inclusive cross section
is shown in Fig.~\ref{fig5} for the  PWIA calculation. The inclusive cross
section
is formed additively by the contributions for $\Lambda n$,
$\Sigma^0 n$ and $\Sigma^-p$ production. Above 
the $K^+\Sigma N$ threshold the $\Lambda n$ contribution becomes smaller
while the two parts for $\Sigma^0n$ and $\Sigma^-p$ production
contribute about equally. Note that the $\Sigma^-p$ contribution
 results from the elementary $K^+$ production on the neutron.

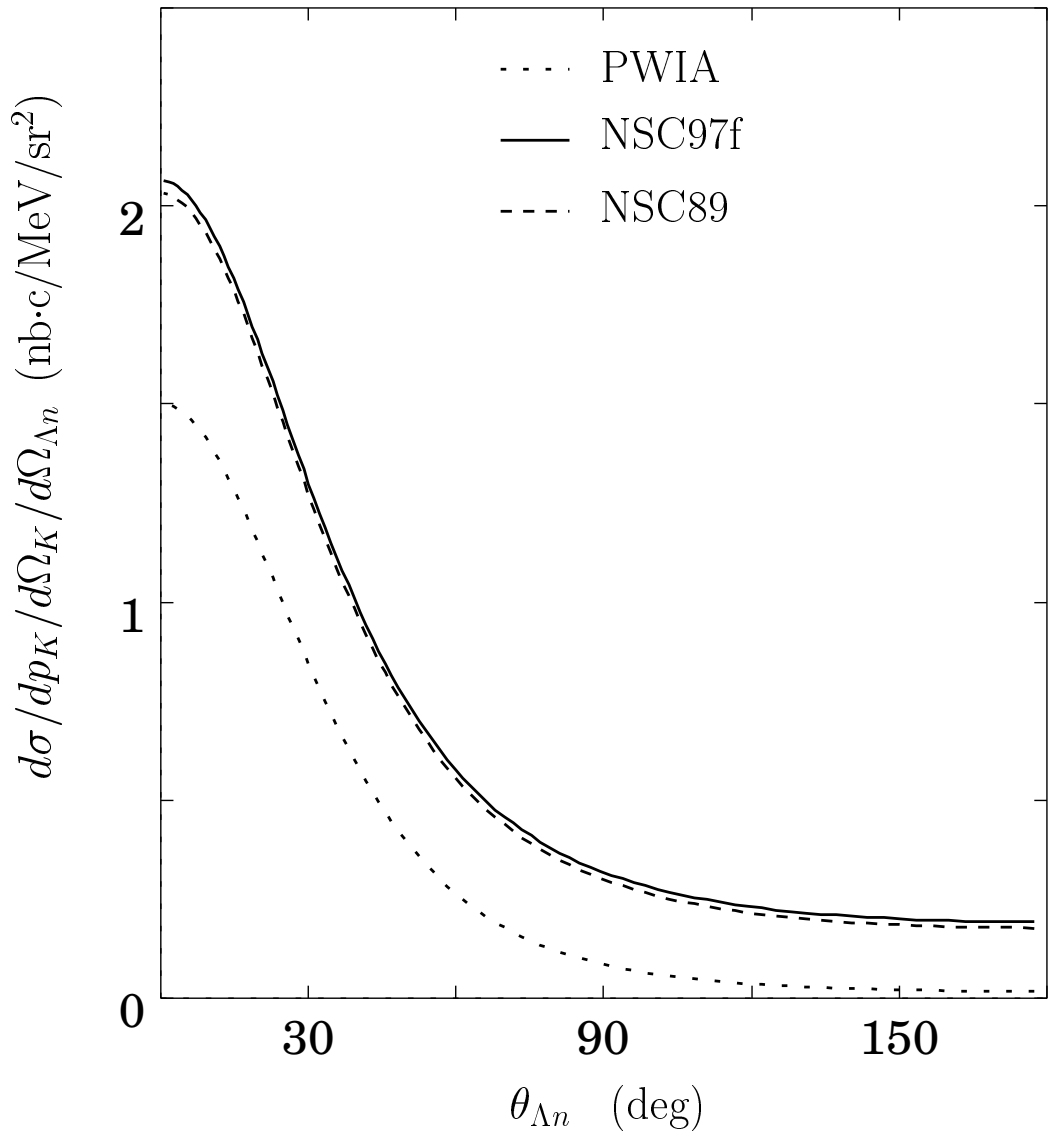
\begin{figure}
\caption{
The exclusive cross section
$\gamma (d,K^+\Lambda )n$ cross section for
 $\theta_K=0^\circ$, lab momentum $p_K=972$ MeV/$c$  and 
 photon lab energy $E_\gamma=1.3$ GeV
as a function of the $\Lambda$ scattering angle in the 
$\Lambda N$ c.m. system. The plane wave prediction is compared
 to two $YN$ force calculations.
\label{fig6}}
\end{figure}

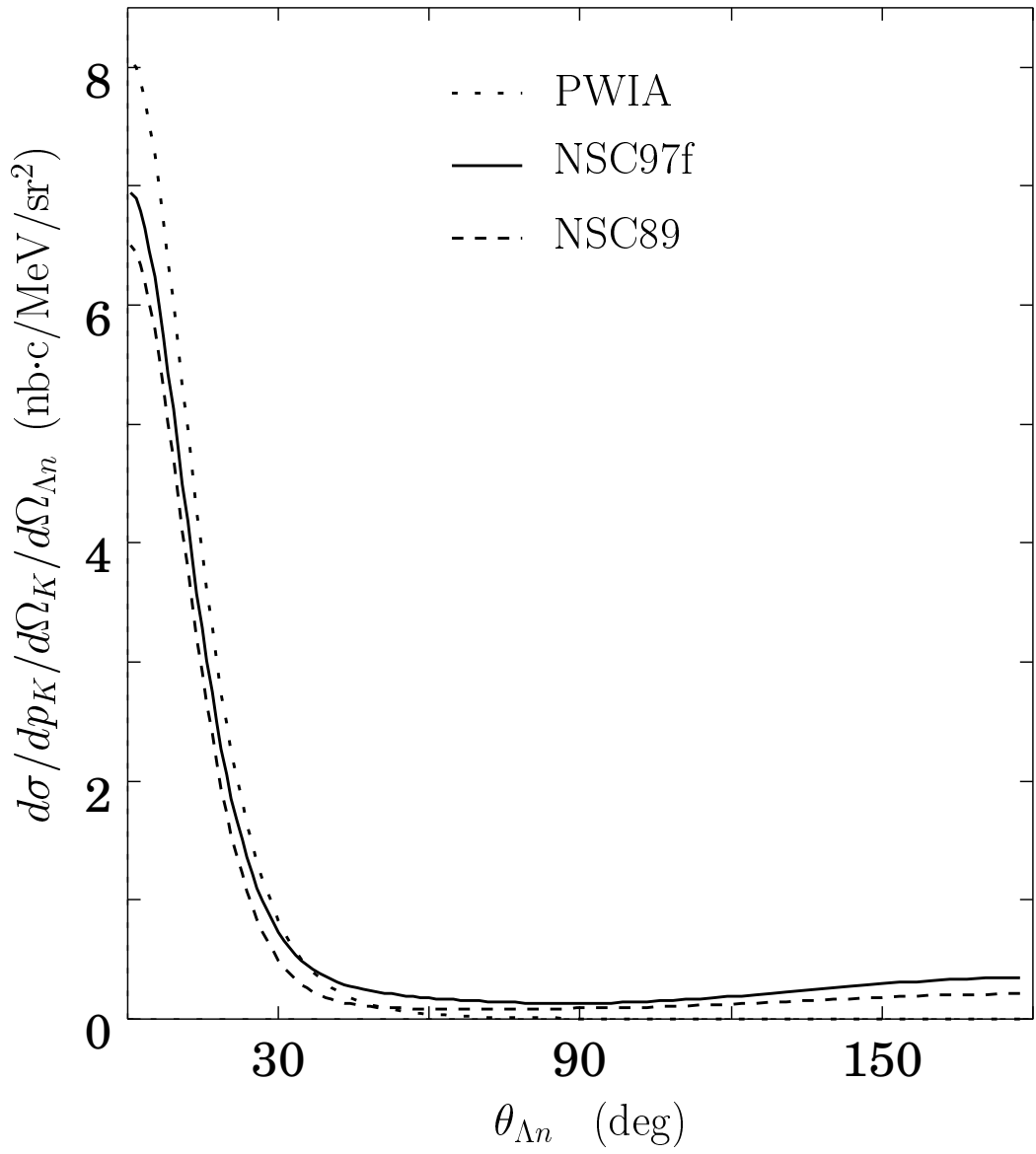
\begin{figure}
\caption{
The same as in Fig.~\ref{fig6} for $p_K=870$ MeV/c. The FSI effects
scatter the 
$\Lambda$'s to larger angles than in PWIA.
\label{fig7}}
\end{figure}

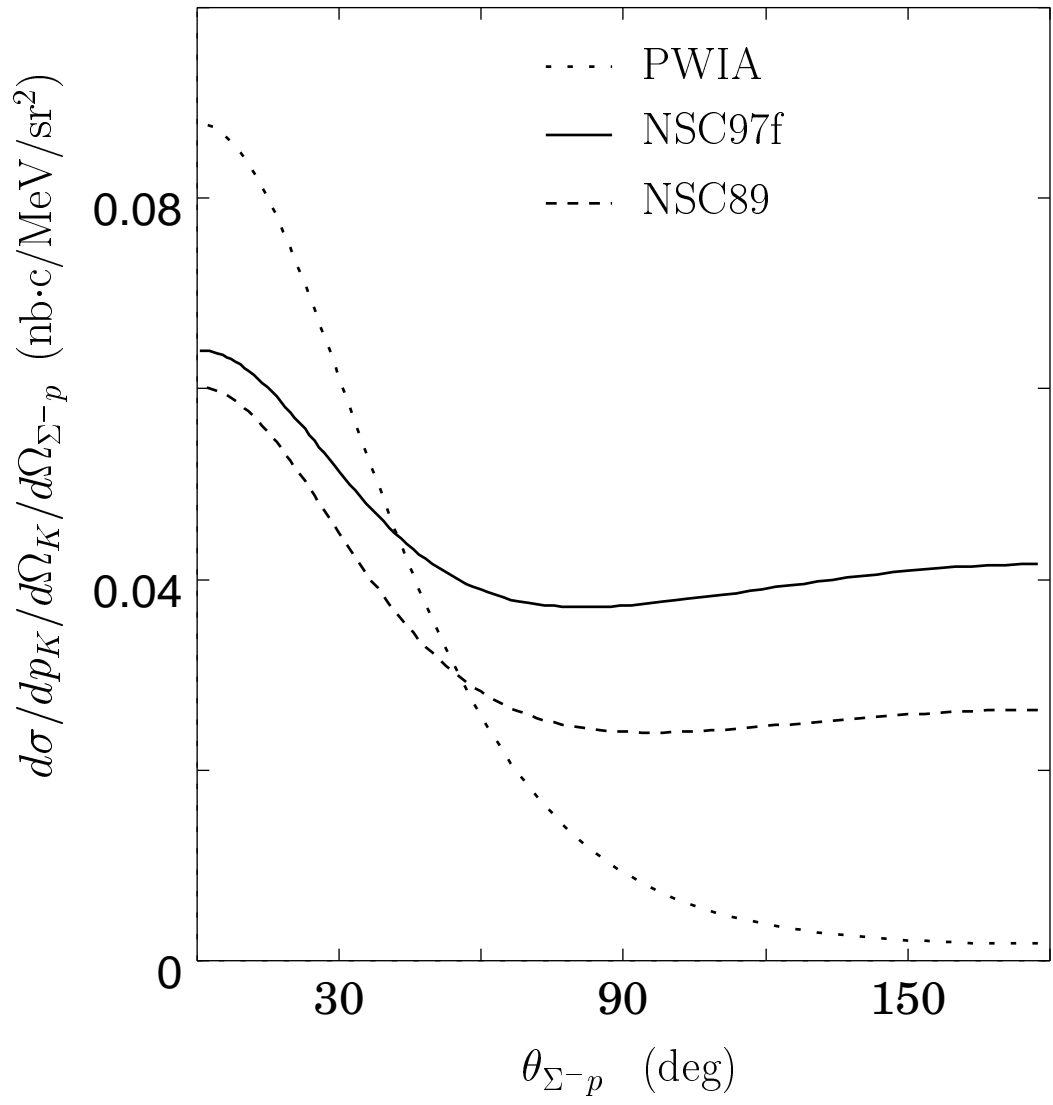
\begin{figure}
\caption{
The exclusive cross section
$\gamma (d,K^+\Sigma^- )p$ cross section for
 $\theta_K=0^\circ$, lab momentum $p_K=865$ MeV/$c$  and 
 photon lab energy $E_\gamma=1.3$ GeV
as a function of the $\Sigma^-$ scattering angle in the 
$\Sigma^-p$ c.m. system. The plane wave prediction is compared
 to two $YN$ force calculations which are strikingly different.
\label{fig8}}
\end{figure}

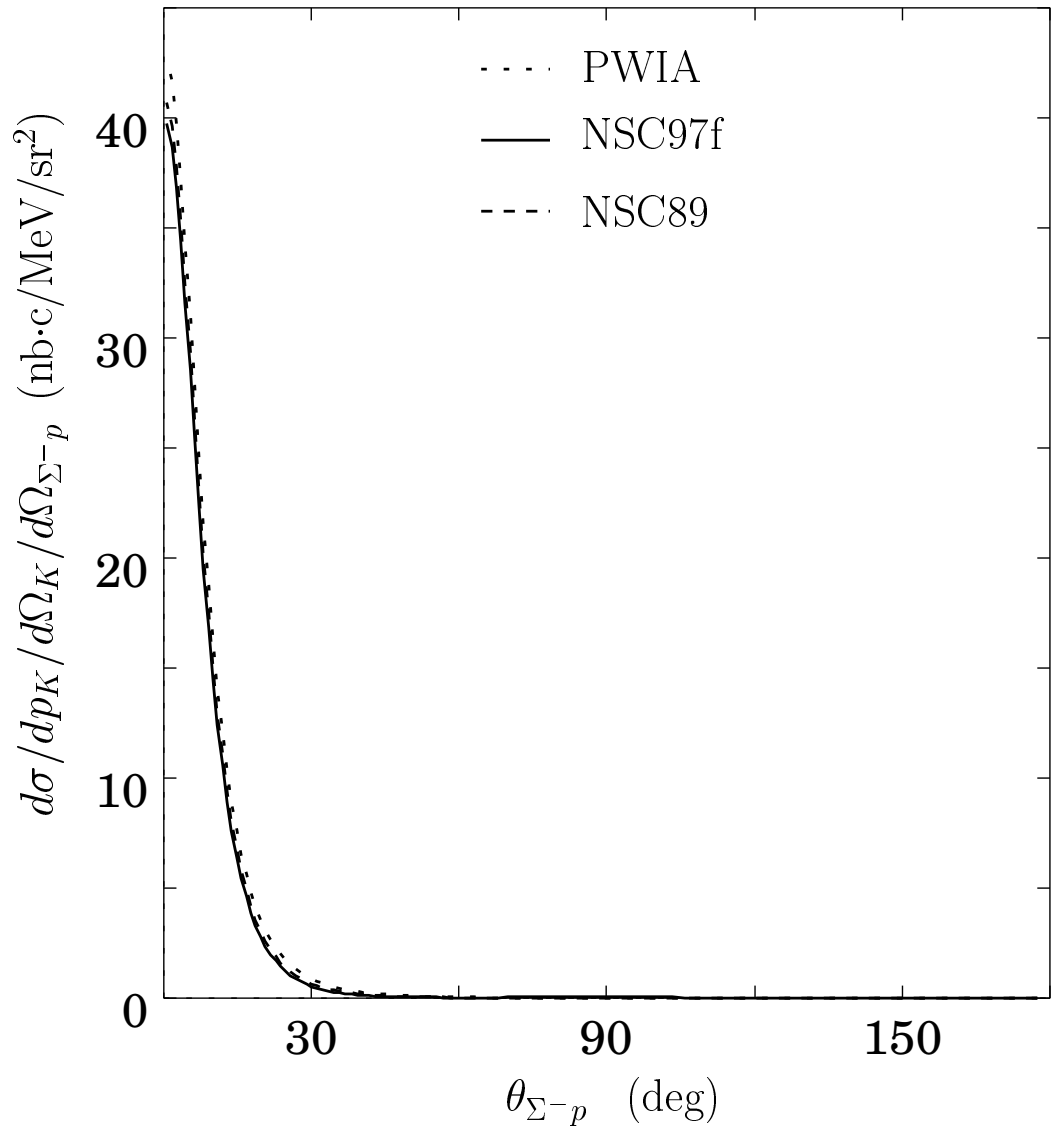
\begin{figure}
\caption{
The same as in Fig.~\ref{fig8} for $p_K=810$ MeV/$c$.
\label{fig9}}
\end{figure}

There are many options to display the information contained in the
exclusive
cross section; we show in Figs.~\ref{fig6}$-$\ref{fig9}
angular distributions of the
hyperons in the hyperon-nucleon c.m. frame for the $K^+$ meson
emerging in the direction of the photon ($\theta_K =0^\circ$). For
$K^+ \Lambda$ production
we have chosen two $K$-meson momenta,
 $p_K=972$ MeV/c, shown in Fig.~\ref{fig6}, and $p_K=870$ MeV/c,
the region just below the $K^+\Sigma N$ threshold (Fig.~\ref{fig7}). 
In both cases FSI effects are found to be significant.
As shown in Fig.~\ref{fig6}, the calculations including FSI lie 
considerably above the PWIA results. The difference between 
the two FSI results is small at this kinematics.  
Fig.~\ref{fig7} displays cross sections at
 $p_K=870$ MeV/c, here the effect of FSI is to scatter $\Lambda$'s
 to larger angles compared to PWIA. At very backward angles
 the PWIA result is basically zero while the FSI calculations still show 
 some strength.
 
In the case of $\Sigma^- p$ production we also present results at 
two different kaon momenta, one just
above the $\Sigma$ threshold and one in the peak region of the
inclusive cross section. Fig.~\ref{fig8} shows dramatic differences
not only between the PWIA and the FSI results but also between
the calculations that employ the NSC97f and the NSC89 forces.
For backward angles, the results differ by more than 50\%.
Measurements should easily be able to distinguish between these
 possibilities.
In contrast, no differences are seen in Fig.~\ref{fig9}, illustrating how
 important
it is to choose the proper kinematics.

Further observables for the exclusive process, like energy distributions
for fixed $K^+$ meson and hyperon angles, will be studied in a forthcoming
article.
Especially configurations where the relative energy between the hyperon
 and nucleon goes to zero might be of interest, since this
might allow extracting hyperon-nucleon scattering lengths. The poles
in the hyperon-nucleon $t$-matrix near the thresholds 
will lead to enhancements, though 
not as spectacular as in the $NN$ case\cite{physrep}. 
Other quantities that will be studied include
polarization observables which are especially accessible 
due to the self-analyzing property of the lambda. Thus,
the lambda recoil polarization in combination with either
linearly or circularly polarized photons could be measured at
Jefferson Lab. The first measurements of kaon photoproduction
on the deuteron are scheduled to take place later this year\cite{mecking}.

\section{Conclusion}
\label{conclusion}

We have evaluated the  
  $\gamma( d,K^+)$
inclusive and the 
 $\gamma( d,K^+ Y)$ 
exclusive processes using the modern $YN$ interactions
NSC89 and NSC97f.
Both include $\Lambda$-$\Sigma$ conversion and give 
the correct $^3_\Lambda$H binding energy.
The deuteron wavefunction was based on the 
Nijmegen93 $NN$ potential.
The aim was to search for final state interaction effects in 
the $\Lambda N$ and $\Sigma N$ systems. In the inclusive
 cross section  we found effects near the $K^+\Lambda N$ 
 and the $K^+\Sigma N$ thresholds which might be measurable.
Especially for the latter case the two $YN$ potential
predictions are quite different, reflecting the different
underlying $S$-matrix pole structure for the two $YN$ forces.
The exclusive process shows significant
FSI effects which we displayed for the angular distributions
of the hyperons.
Around the $K^+\Sigma N$ threshold FSI effects
are especially interesting and the two $YN$ forces show quite 
different effects. Future data should easily be able to distinguish
between the different $YN$ forces.

\acknowledgements
W.G. would like to
thank the Deutsche Forschungsgemeinschaft for financial support and
the Okayama
University of Science for the warm hospitality.
K.M. thanks the few-body group of the Ruhr University Bochum for 
the kind hospitality during his stay. C.B. acknowledges the 
support from DOE grant DE-FG02-95ER-40907. T.M. thanks 
the Okayama University of Science for the hospitality and 
acknowledges the University Research for Graduate Education (URGE) grant.


\pagebreak
\noindent

\noindent

\noindent

\noindent

\noindent

\noindent

\noindent

\noindent

\noindent

\noindent

\end{document}